\documentclass[12pt,preprint]{aastex}

\newcommand{\prt}{\partial}

\begin{document}

\title{Large scale properties in turbulent spherically
symmetric accretion}

\author{Arnab K. Ray}
\affil{Harish--Chandra Research Institute, \\Chhatnag Road, Jhunsi,
Allahabad 211 019, INDIA}
\email{arnab@mri.ernet.in}

\and

\author{J. K. Bhattacharjee}
\affil{Department of Theoretical Physics, \\ Indian Association for the
Cultivation of Science, \\Jadavpur, Calcutta (Kolkata) 700 032, INDIA}
\email{tpjkb@mahendra.iacs.res.in}

\begin{abstract}
The role of turbulence in a spherically symmetric accreting system has
been studied on very large spatial scales of the system. This is also a 
highly subsonic flow region and here the accreting fluid has been treated 
as nearly incompressible. It has been shown here that the coupling of the 
mean flow and the turbulent fluctuations, gives rise to a scaling relation 
for an effective ``turbulent viscosity". This in turn leads to a dynamic 
scaling for sound propagation in the accretion process. As a consequence 
of this scaling, the sonic horizon of the transonic inflow solution is 
shifted inwards, in comparison with the inviscid flow. 
\end{abstract}


\section{Introduction}

The purpose of this work has been to study the dynamic scaling 
behaviour of the coefficients of viscosity arising out of turbulence 
in a spherically symmetric accreting system, and how such scaling 
behaviour leads to a scale dependence for the speed of sound as well. 
In its turn this will be shown to have an important bearing on the 
sonic point of the transonic inflow solution, since it is with the 
speed of sound that both the bulk velocity of the flow and the sonic 
point are scaled. A study of this kind should be useful in addressing
recent observational discrepancies, for which the classical (and 
inviscid) Bondi theory has proved somewhat inadequate. 

In accretion studies, turbulence is of great relevance, since in
almost all cases of physical interest, the accreting astrophysical 
fluid is in a turbulent state \citep{arc99}. It is presently a well 
established
fact that in the two dimensional case of a thin accretion disc, 
viscous shearing between two differentially rotating adjacent
layers, accomplishes the outward transport of angular momentum
and effectively facilitates the infall of matter \citep{pri81,fkr92}.
However, in this situation, ordinary molecular viscosity has been
known to be quite an inadequate mechanism to explain the rate of the 
transport process. On the other hand,
it is to a very high value of the Reynold's number that the flow 
corresponds, and as such the flow is widely acknowledged to be turbulent
\citep{fkr92,bh98}.
In such a situation, turbulence --- as 
quantitatively characterized by a ``turbulent viscosity"
in the Navier-Stokes' equation --- becomes a prime candidate for a
physical mechanism that brings about an enhanced outward transport of 
angular momentum. The very well known $\alpha$ prescription of \citet{ss73}
is based on this principle. 

As opposed to the facilitating role that it plays in a rotationally
accreting flow, viscosity --- even presumably ``turbulent viscosity" ---
affects the more paradigmatic spherically symmetric accreting flow,
somewhat differently. In the latter case, it has been seen that the
role of viscosity is actually directed towards inhibiting the process
of gravity driven infall, and in doing so, viscosity also sets up 
a limiting length scale for the effectiveness of gravity --- the ``viscous 
shielding radius" \citep{ray03}.
However, as in the thin disc system, even in the spherically
symmetric case, molecular viscosity would be far too weak a mechanism
to bring about a significant quantitative impact. It would then be 
well worth investigating into the question of how significantly would 
spherically symmetric accretion be affected by a large and scale dependent
turbulent viscosity. Following the
qualitative insights obtained with the introduction of molecular
viscosity in the governing hydrodynamical equations, it is possible 
as a matter of standard practice to study both the qualitative and the 
quantitative extent of the influence of an {\em effective} turbulent 
viscosity on the hydrodynamical processes. 
The main purpose here would be to
show that the turbulent fluctuations of the interstellar medium are capable 
of renormalizing on large length scales, the 
small molecular viscosity given in the Navier-Stokes equation. This
renormalized effective viscosity, as pictured by \citet{hei48} in his
theory of turbulence, can very well be instrumental in setting a 
noticeable limiting length scale on the effectiveness of gravity to drive
the accretion process. Indeed, the renormalizing of the viscosity would
be robust enough to make viscosity be comparable with pressure, 
on the same scale of length. 

Related to this contention, one 
example that may be cited is that of the resistive role against gravity 
that turbulence plays in another spherically symmetric 
system of astrophysical interest --- that of the self-gravity
driven Jeans collapse of a gas cloud without angular momentum. Studies 
carried out by 
\citet{bona87a,bona92b} have shown that turbulence acts as a stabilizing
agent against a self-gravity driven collapse. A renormalization approach
has shown that a renormalized turbulent pressure acts against gravity. In 
addition to this, what is being contended in this work is that a scaled-up 
renormalized viscosity also enfeebles the influence of gravity. Both the 
pressure term and the viscous term derive from the stress tensor in the 
Navier-Stokes equation \citep{tl72}. For the spherically symmetric 
case, the contribution comes only from one diagonal (hence, isotropic)
element of the stress tensor. In such a situation, the renormalized
pressure and the viscosity terms would both manifest themselves through
the same physical effect, and the physical contents of the arguments
presented by \citet{bona87a,bona92b} for the self-gravity driven 
spherical collapse of a gas cloud, would match those in this study of 
turbulent spherically symmetric accretion. Having noted this point, it 
would also be instructive to have an understanding of the difference 
between the two physical cases being compared here. Whereas \citet{bona92b}
have studied the response to large scale density perturbations on a 
stationary turbulent solution in a self-gravity driven system, what is
being studied in 
this work, is the influence of spontaneous fluctuations on 
the mean stationary solution of a system, in which gravity comes into play
through an external accretor. 
An analysis of the latter nature is all the more contextual with regard to
accretion studies, because spherically symmetric accretion is exemplified
by the infall of interstellar matter on to an isolated accretor, and
it has been well recognized that the interstellar medium displays
turbulent behaviour \citep{jl69,jls69,lj76}. 

While dwelling on this matter, it would be important to mention that
certain previous studies in spherically symmetric accretion on to a 
black hole, have in fact quantitatively accounted for the physical role 
of turbulence in very efficiently converting gravitational energy to 
radiation. In the works of \citet{mes75} and \citet{ms77}, it has been argued 
that for spherical
accretion on to a massive black hole, turbulent dissipation would be one 
of the factors which would result in the luminosity of the system being 
enhanced by quite a few orders of magnitude --- indeed to such an extent
that the spherically symmetric system could be compared with disc models
as an X-ray source. 

\section{The equations of turbulent spherical accretion}

The effect of turbulent fluctuations has been studied here on very 
large spatial (and therefore highly subsonic scales) of the spherically 
symmetric accreting system. The physical effects of turbulence 
are appreciably manifested on these scales. Since all physically 
feasible flow solutions have to pass through the subsonic flow region,
the turbulent fluctuations here must have a significant influence on
the flow. And more to the point, on these subsonic scales, the flow
could be studied in the nearly incompressible regime. 

Turbulence is an attribute of the fluid flow \citep{tl72}, while molecular 
viscosity is an intrinsic physical property of the fluid. And yet the 
two can be very closely related to each other through the Navier-Stokes 
equation \citep{fri99}, which, as one of the governing equations of the 
flow, is given by 
\begin{equation}
\label{nv}
\frac{\prt {\bf{v}}}{\prt t} +({\bf{v}}{\bf{\cdot}}{\bf{\nabla}}){\bf{v}}
+ \frac{{\bf{\nabla}} P}{\rho} + \frac{GM}{r^2} \hat{\bf r}= \nu 
{\bf{\nabla}}^2 {\bf{v}}+ 
\mu {\bf{\nabla}}({\bf{\nabla}}{\bf{\cdot}}{\bf{v}})
\end{equation}
where $\nu$ and $\mu$ are the two kinematic coefficients of viscosity.
The pressure $P$ is related to the density through a general
polytropic equation of state $P=k{\rho}^{\gamma}$. Here $\gamma$ is the 
polytropic exponent with an admissible range given by $1<{\gamma}<5/3$ 
--- these restrictions having been imposed by the isothermal limit and the 
adiabatic limit respectively. The flow is also governed by the continuity 
equation, which is given by 
\begin{equation}
\label{con}
\frac{\prt \rho}{\prt t}+{\bf{\nabla}}{\bf{\cdot}}({\rho \bf{v}})=0
\end{equation}

The total velocity and density fields are written as 
${\bf{v}} = {\bf{v_0}} + {\bf{u}}$ and $\rho = \rho_0 + \delta \rho $, 
in which $\bf{v_0}$ and $\rho_0$, which are functions of the radial 
coordinate only, are the mean velocity and density profiles for the 
spherically symmetric transonic flow, while $\bf u$ and $\delta \rho$ 
generally are time-dependent and three-dimensional random fluctuations 
about the transonic solution. The implicit understanding here is that 
in an accreting system naturally evolving in real time, the transonic 
solution is accorded primacy over all possible other stationary solutions
in both the inviscid \citep{bon52,gar79,rb02} and viscous regimes 
\citep{an67}. Under the assumption that cross-correlations of the 
density and the velocity fluctuations would be negligible, i.e. 
$\langle {\bf{\nabla}}{\bf{\cdot}}({\bf{u}} \delta \rho) \rangle = 0$, 
the average (and steady) solutions would be obtained as  
\begin{equation}
\label{steadcon}
{\bf{\nabla}}{\bf{\cdot}}(\rho_0 {\bf{v_0}}) = 0
\end{equation}
and
\begin{equation}
\label{steadnv}
({\bf{v_0}}{\bf{\cdot}}{\bf{\nabla}}){\bf{v_0}} 
+ \langle ({\bf{u}}{\bf{\cdot}}{\bf{\nabla}}){\bf{u}} \rangle
+ {\frac{\gamma k}{\gamma -1}}
{\bf{\nabla}}\rho_0^{\gamma -1} + {\frac{GM}{r^2}}\hat{\bf{r}} =
\nu {\bf{\nabla}}^2 {\bf{v_0}} + \mu {\bf{\nabla}}({\bf{\nabla}}
{\bf{\cdot}}{\bf{v_0}})
\end{equation}

In the nearly incompressible regime, only the first order term in 
the expansion of the density fluctuations about the mean density 
need be retained. The fluctuating density and velocity fields are 
therefore seen to satisfy, 
\begin{equation}
\label{flucden}
{\frac{\prt}{\prt t}}{\delta \rho} + {\bf{\nabla}}{\bf{\cdot}}(\rho_0
{\bf{u}}) + {\bf{\nabla}}{\bf{\cdot}}({\bf{v_0}} \delta \rho) +
{\bf{\nabla}}{\bf{\cdot}}({\bf{u}} \delta \rho) =0 
\end{equation}
and
\begin{equation}
\label{flucnv}
{\frac{\prt {\bf{u}}}{\prt t}} +
({\bf{u}}{\bf{\cdot}}{\bf{\nabla}}){\bf{u}} + 
{\bf{\nabla}}\left[c_s^2 \frac{\delta \rho}{\rho_0} \right]
= \nu {\bf{\nabla}}^2 {\bf{u}} 
+ \mu {\bf{\nabla}}({\bf{\nabla}}{\bf{\cdot}}{\bf{u}})
- \Big[({\bf{u}}{\bf{\cdot}}{\bf{\nabla}}){\bf{v_0}}
+ ({\bf{v_0}}{\bf{\cdot}}{\bf{\nabla}}){\bf{u}}
- \langle ({\bf{u}}{\bf{\cdot}}{\bf{\nabla}}){\bf{u}} \rangle \Big]
\end{equation}
respectively, with $c_s$ being the steady value of the speed of sound, 
which is related to the mean density by $c_s^2={\gamma}k \rho_0^{\gamma -1}$.

In the very much subsonic region of the flow, the variation of the
mean density $\rho_0$ may be neglected, since in this region the mean 
density very closely assumes an ambient value, which is a constant. 
Under this approximation, equation (\ref{steadcon}) gives the relation
${\bf{\nabla}}{\bf{\cdot}}{\bf{v_0}} \cong 0$. Furthermore, on these 
scales the continuity equation also governs the asymptotic behaviour 
of the mean velocity, which implies that the variation of the mean 
velocity, at its most rapid, is given by 
$v_0 \sim r^{-2}$ \citep{pso80, skc90}. 
On the other hand (under these asymptotic conditions) the turbulent velocity 
fluctuations are much greater than the mean velocity itself, and in fact are of 
the order of the speed of sound. Hence on these large length scales, ignoring 
all terms involving the mean velocity and the gradient of the mean density and 
its fluctuations, it would be meaningful to retain only the primary signature 
of a compressible flow, namely 
${\bf{\nabla}}{\bf{\cdot}}{\bf{u}} \not = 0$, whence 
equation (\ref{flucden}) simplifies to
\begin{equation}
\label{simfluc}
\frac{1}{\rho_0} \frac{\prt}{\prt t}{\delta \rho} 
+ {\bf{\nabla}}{\bf{\cdot}}{\bf{u}} = 0
\end{equation}
which is an expression that has found quite regular mention in the study of a 
nearly incompressible fluid flow with random fluctuations \citep{syko90,bhat93}.

At this stage it should be important to be assured of the consistency in 
neglecting the higher powers of $\delta \rho/\rho_0$. Under the chosen
working approximations, the terms in the left hand side of 
equation (\ref{flucnv})
can be written as $\dot{u}_\alpha$, $u_\beta \prt_\beta u_\alpha$ and 
$c_s^2 \prt_\alpha (\delta \rho/\rho_0)$ respectively. If $u_\alpha$ and 
$c_s$ are to scale as $L^\epsilon$, then the time $t$ scales as 
$L^{1 - \epsilon}$, while $(\delta \rho/\rho_0)$, of course, remains 
independent of any scaling. Here $\epsilon$ is arbitrary, but anticipating
that a one-loop calculation will yield a positive value for $\epsilon$, all
nonlinearities involving 
$\delta \rho/\rho_0$ (with $\rho_0$ being asymptotically
a constant) may be ignored in favour of $u_\beta \prt_\beta u_\alpha$. 
This, arguably, should suffice for a study of the scaling dependence 
in the flow. If the resulting calculations lead to a positive $\epsilon$, 
the adopted procedure would be justified and would be consistent with 
itself. That this is precisely what happens, will be demonstrated in 
the following sections.
 
Equation (\ref{flucnv}) is likewise simplified, and closed with the help of 
equation (\ref{simfluc}), to give
\begin{equation}
\label{eqinu}
{\frac{\prt {\bf{u}}}{\prt t}} +
({\bf{u}}{\bf{\cdot}}{\bf{\nabla}}){\bf{u}} = 
c_s^2{\bf{\nabla}}\Big({\frac{\prt}{\prt t}}\Big)^{-1}
({\bf{\nabla}}{\bf{\cdot}}{\bf{u}})+ \nu {\bf{\nabla}}^2 {\bf{u}}
+ \mu {\bf{\nabla}}({\bf{\nabla}}{\bf{\cdot}}{\bf{u}}) + {\bf{f}}
\end{equation}
in which ${\bf{f}}= - \left[({\bf{u}}{\bf{\cdot}}{\bf{\nabla}}){\bf{v_0}}
+({\bf{v_0}}{\bf{\cdot}}{\bf{\nabla}}){\bf{u}} - \langle ({\bf{u}}{\bf{\cdot}}
{\bf{\nabla}}){\bf{u}} \rangle \right]$.  
The primary complication at this stage is in this term ${\bf{f}}$,
which couples the fluctuating flow to the mean flow. Gravity of the 
central accretor maintains the mean flow, from which energy
is transferred to the fluctuating flow, through its coupling with the
mean flow. So effectively what happens is that the turbulent fluctuations
are sustained by gravitation, via the nonlinear coupling in the term 
${\bf{f}}$. Various approximations in the theory of turbulence have 
involved a modelling of this nature of energy input to the turbulent 
flow. Prandtl's mixing length theory is one of the most well known
\citep{fab95,arc99}. A more recent point of view treats this force as an as 
yet unspecified force external to the turbulent flow \citep{fns77,ddm79}. 
Its dependence on the random field ${\bf{u}}$, makes it random and hence the 
modelling endows ${\bf{f}}$ with random properties. Even for this accretion 
problem it would therefore be quite possible to conceive of a randomly forced 
turbulent flow described by (for the nearly incompressible flow that is being 
studied here)
\begin{equation}
\label{teneqinu}
{\prt}_t u_i + (u_j {\prt}_j)u_i = c_s^2 {\prt}_i 
({\prt}_t^{-1} {\prt}_j u_j ) + \nu {\prt}_j {\prt}_j u_i
+ \mu {\prt}_i ({\prt}_j u_j) + f_i
\end{equation} 
in which, for the Gaussian forcing, the correlation function is specified as
\begin{equation}
\label{corr}
\langle f_i ({\bf{r}},t) f_j ({\bf{r}},t) \rangle = {\delta}_{ij}
C_0 \big( \vert {\bf{r}} - {\bf{r}^{\prime}} \vert \big)
{\delta} \big(t - t^{\prime} \big)
\end{equation}
These two equations (\ref{teneqinu}) and (\ref{corr}) will be necessary to 
develop a dynamic scaling theory for the turbulent spherically symmetric flow.

\section{Dynamic scaling for turbulent spherical accretion}

To carry out a dynamic scaling analysis with the help of equations 
(\ref{teneqinu}) and (\ref{corr}), it would be 
convenient to work in Fourier transform space. 
This would necessitate writing 
\begin{equation}
\label{foutran}
u_i ({\bf{r}},t) = {\frac{1}{(2\pi)^2}} {\int} u_i ({\bf{k}}, \omega)
e^{i({\bf{k}}{\bf{\cdot}}{\bf{r}} - \omega t)} {\mathrm{d}}^3 {\bf{k}} \, 
{\mathrm{d}} \omega
\end{equation}
in terms of which equation (\ref{teneqinu}) becomes 
\begin{equation}
\label{traneqinu}
\Big[ (- i \omega + \nu k^2) {\delta}_{ij} + \mu k_i k_j \Big] u_j
- c_s^2 {\frac{k_i k_j}{i \omega}} u_j = f_i - i \sum_{{\bf{p}},
{\omega}^{\prime}} p_j u_j ({\bf{k}} - {\bf{p}}, {\omega}^{\prime})
u_i ({\bf{p}}, \omega - {\omega}^{\prime})
\end{equation}

The technique that has been adopted here is to expand the velocity 
field as $u_i = u_i^{(0)} + u_i^{(1)} + u_i^{(2)} + {\ldots} \,$, 
in which $u_i^{(0)}$ is the solution in the absence of the nonlinear 
term. The subsequent terms are the effect of the nonlinear term in 
equation (\ref{traneqinu}). The lowest order solution can then be written as 
\begin{equation}
\label{lowodd}
u_i^{(0)} = G_{ij}^{(0)} f_j 
\end{equation} 
in which
\begin{equation}
\label{greeninv}
\Big[G_{ij}^{(0)}\Big]^{-1} = \Big(- i \omega + \nu k^2 \Big) 
\delta_{ij} - \Big({\frac{c_s^2}{i \omega}} - \mu \Big) k_i k_j
\end{equation}
The first-order correction, $u_i^{(1)}$, satisfies 
\begin{equation}
\label{first}
\Big[\Big(- i \omega + \nu k^2 \Big) \delta_{ij}
- \Big({\frac{c_s^2}{i \omega}} - \mu \Big) k_i k_j \Big]
u_i^{(1)} = - i \sum_{{\bf{p}},{\omega}^{\prime}} p_j u_j^{(0)} 
({\bf{k}} - {\bf{p}}, \omega^{\prime})
u_i^{(0)} ({\bf{p}}, \omega - \omega^{\prime})
\end{equation}
and its solution is given by
\begin{equation}
\label{solfirst}
u_i^{(1)} = -i G_{ij}^{(0)}({\bf{k}}, \omega)
\sum_{{\bf{p}},{\omega}^{\prime}} p_k u_k^{(0)} 
({\bf{k}} - {\bf{p}}, \omega^{\prime})
u_j^{(0)} ({\bf{p}}, \omega - \omega^{\prime})
\end{equation}

As has been stressed by \citet{hei48}, the momentum transfer term, given 
in the right hand side of equation (\ref{traneqinu}), gives rise 
to an effective turbulent 
shear viscosity (the eddy viscosity) for an incompressible flow. 
This is the physical content of all subsequent theories --- the different
kinds of renormalized perturbation expansion \citep{mc90}, the 
renormalization group \citep{mc90},
the self-consistent mode coupling \citep{mc90} and the very recent 
Lagrangian picture approach \citep{lp95a,lp95b}. 
In this compressible case, it will be easy to see that the 
right hand side of equation (\ref{traneqinu}) will give rise to 
the effective shear viscosity,
bulk viscosity and speed of sound. The simplest way of arriving at this
result is to examine the average value of the right hand 
side of equation (\ref{traneqinu}),
averaged over the distribution of the random force $f_i$. This is done
perturbatively. In what follows in this section, the 
salient results of the perturbative analysis have been brought forth. 
The details of the 
calculations have been presented in the Appendix. 

The averaging of the nonlinear term in equation (\ref{traneqinu}) will 
lead to its equivalent linearized representation (see the Appendix), 
given by 
\begin{equation}
\label{avnonlin}
\langle  - i \sum_{{\bf{p}}, \omega^{\prime}} 
p_j u_j ({\bf{k}} - {\bf{p}},{\omega}^{\prime})
u_i ({\bf{p}}, \omega - \omega^{\prime}) \rangle \equiv 
- \sigma_{il}^{(0)} ({\bf{k}}, \omega ) u_l^{(0)} ({\bf{k}}, \omega )
\end{equation}
where
\begin{equation}
\label{sigmaexp}
\sigma_{il}^{(0)} ({\bf{k}}, \omega )
= 2 \sum_{{\bf{p}},{\omega}^{\prime}} p_j k_k \Big[G_{jl}^{(0)}
({\bf{k}}-{\bf{p}},{\omega}^{\prime}) \tilde{C}_{ki}^{(0)}({\bf{p}},
\omega -  \omega^{\prime}) + G_{il}^{(0)}
({\bf{p}}, \omega - \omega^{\prime}) \tilde{C}_{kj}^{(0)}({\bf{k}}
- {\bf{p}}, \omega^{\prime}) \Big]
\end{equation}
and in which
\begin{equation}
\label{ceecap}
\tilde{C}_{kj}^{(0)}({\bf{p}}, \omega^{\prime}) = 
G_{kl}^{(0)}({\bf{p}}, \omega^{\prime})C_0({\bf{p}})
G_{lj}^{(0)}(-{\bf{p}},- \omega^{\prime})
\end{equation}

Clearly, $\tilde{C}_{ij}^{(0)}$ is the zeroth-order correlation
function for the velocity field. The nonlinear term in the equation of
motion now has the structure $\sigma_{il}^{(0)} ({\bf{k}}, \omega ) 
u_l^{(0)} ({\bf{k}}, \omega )$ in this lowest order of the perturbation
theory. The coefficient $\sigma_{il}^{(0)}$ can clearly be identified
as making a contribution to the two coefficients of viscosity and the 
speed of sound, by comparing with the linear term in the equation of 
motion shown in equation (\ref{traneqinu}). 
The $\sigma_{il}^{(0)}$ that has been obtained is called 
the self energy and constitutes a dressing of the bare coefficients.
This is exactly in conformity with all the different ways of doing 
perturbation theory \citep{mc90}.
The step beyond perturbation theory goes to say
that as the higher order terms are considered, $\sigma_{il}^{(0)}$ 
will be converted to the full self energy $\sigma_{il}$. It must be
emphasized here that conversion to the full self energy will not 
differently affect the scaling arguments that will be developed here
on the basis of the lowest order in the perturbation theory. 

To make any further progress, it would be instructive to examine the 
structure of 
$\sigma_{ij} ^{(0)} ({\bf{k}}, \omega )u_l^{(0)} ({\bf{k}}, \omega )$. 
By comparing with the form of $\big[{G_{ij}^{(0)}}\big]^{-1}$ in 
equation (\ref{greeninv}), it is possible to write 
\begin{eqnarray}
\label{compsig}
\sigma_{ij}^{(0)} ({\bf{k}}, \omega )
&=& 2 \sum_{{\bf{p}}, \omega^{\prime}} p_m k_n \Big[G_{mj}^{(0)}
({\bf{k}}-{\bf{p}}, \omega^{\prime}) \tilde{C}_{ni}^{(0)}({\bf{p}},
\omega -  \omega^{\prime})\nonumber \\ 
& & \qquad \qquad + G_{im}^{(0)}
({\bf{p}}, \omega -  \omega^{\prime}) \tilde{C}_{jn}^{(0)}({\bf{k}}
- {\bf{p}}, \omega^{\prime}) \Big]\nonumber \\
&=& k^2 \Big[\sigma_1^{(0)}({\bf{k}, \omega})\delta_{ij} 
+\sigma_2^{(0)}({\bf{k}}, \omega) \frac{k_i k_j}{k^2} \Big]
\end{eqnarray}
where $\sigma_1^{(0)}$ and $\sigma_2^{(0)}$ are the frequency and 
momentum dependent components of the self energy tensor which must have
the structure shown in equation (\ref{compsig}) from 
the isotropy of space. Evidently,  
$\sigma_1^{(0)}({\bf{k}}, \omega = 0)$ dresses the shear viscosity, while
$\sigma_2^{(0)}({\bf{k}}, \omega)$ dresses the bulk viscosity and the
speed of sound. To have any information about the dressing of the speed 
of sound, the $(i \omega)^{-1}$ part would have to be extracted from 
$\sigma_2^{(0)}({\bf{k}}, \omega)$ and the rest of the integral would
have to be evaluated at $\omega = 0$, to yield the dressed bulk viscosity. 

The renormalization of $\nu$, $\mu$ and $c_{s}$ converts them into the
renormalized quantitites $\tilde{\nu}$, $\tilde{\mu}$ and $\tilde{c}_s$ 
respectively. To have any idea of how the two coefficients of viscosity 
and the speed of sound get renormalized, the Green's function would have 
to be written out by inversion of the matrix implied by 
$\big[G_{ij}^{(0)}\big]^{-1}$ in equation (\ref{greeninv}). Substitution 
of the unrenormalized quantities in the Green's function by the 
renormalized ones (see the Appendix), will then give the fully dressed 
Green's function as
\begin{equation}
\label{dresgreen}
G_{ij}({\bf{k}}, \omega) = \frac{1}{-i \omega + \tilde{\nu} k^2}
\Bigg[{\delta}_{ij} - k_i k_j{\frac{\big(\tilde{\mu} - \tilde{c}_s^2/
{i \omega} \big)}{-i \omega + \tilde{\nu} k^2 + k^2  \big(\tilde{\mu} 
- \tilde{c}_s^2/{i \omega} \big)}} \Bigg]
\end{equation} 

The poles of the Green's function, occurring at $\omega = -i \tilde{\nu} k^2$ 
and the roots of $\omega^2 = \tilde{c}_s^2 k^2 - i \omega k^2 (\tilde{\nu}
+ \tilde{\mu})$, would deliver a dispersion relation for $\omega$. It
is satisfying to note here that the second relation (the quadratic in
$\omega$) is identical to the one obtained by \citet{bona92b}, barring
a term arising from the self-gravity of the system that they were
studying. In the long wavelength limit ($k$ small), which contains the
interesting features about the scaling behaviour, the quadratic in
$\omega$ can be approximated as 
\begin{equation}
\label{disp}
\omega \cong \pm \tilde{c}_s k - \frac{i}{2} k^2 (\tilde{\nu} + \tilde{\mu})
\end{equation}

The renormalized quantities will have a power law dependence on $k$. If
dynamic scaling is to be invoked, then the frequency will be proportional 
to some definite power of $k$, which means that all the terms in the right 
hand side of equation (\ref{disp}) must scale in the same way. In physical 
terms this would mean that both the propagating term and the dissipative 
term in equation (\ref{disp}), would be comparably effective on the same 
scale. If a power law were to be written in the form 
$\tilde{\nu} \propto k^{-y}$, then it will clearly also indicate that 
$\tilde{\mu} \propto k^{-y}$ and $\tilde{c}_s \propto k^{1 -y}$. The 
main concern will be to set a value for $y$. To make any progress in that 
direction, the forcing term would have to be specified, which actually would 
imply specifying the correlation function 
$C_0 \big( \vert {\bf{r}} - {\bf{r}^{\prime}} \vert \big)$
in equation (\ref{corr}). Since this term dominates at large distance, a 
scaling form $C_0 \big( \vert {\bf{r}} - {\bf{r}^{\prime}} \vert \big)
\propto {\vert {\bf{r}} - {\bf{r}^{\prime}} \vert}^{\alpha}$ may be 
assumed. This would then transform in the momentum space as 
$C_0 (k) \sim k^{-(D+ \alpha)}$,
where $D$ is the dimensionality of the space. At this point a significant 
departure from \citet{bona92b} is being made, by suggesting that both the 
renormalized $\sigma_1^{(0)}$ and $\sigma_2^{(0)}$, as given by 
equation (\ref{compsig}), would have to be treated as comparable with each 
other. Consequently, information on the scaling of the sound velocity 
(arising from the pressure term) can be had from the scaling of the shear 
viscosity, which is clearly a dissipative effect. 

The fully dressed self-consistent form of equation (\ref{compsig}) 
can now be written down as an integral (see the Appendix), given by 
\begin{equation}
\label{integ}
\sigma_{ij}({\bf{k}}, \omega) = 4 \int \frac{\mathrm{d}^D p}{(2\pi)^D}
\frac{\mathrm{d} \omega^{\prime}}{2\pi} p_m k_n G_{mj}({\bf{k}} - {\bf{p}},
\omega - \omega^{\prime})\tilde{C}_{ni}({\bf{p}}, \omega^{\prime})
\end{equation}
where $\tilde{C}_{ni} = G_{nr} C_0 G_{ri}^{\ast}$. 

For an incompressible flow, the Kolmogorov spectrum requires that 
$\alpha = 0$, to characterize the nature of the transfer of energy between
the mean flow and the fluctuating flow. Since this transfer characteristic
should be independent of the speed of sound, it would be possible to 
write $\alpha = 0$ for the case of near incompressibility being 
discussed here. The left hand side of equation (\ref{integ}) then 
scales as $k^{2-y}$, while the right hand side scales as $k^{2y -2}$. 
For the scaling properties of both sides of equation (\ref{integ}) to agree, 
it should be necessary to set $2-y=2y-2$, which will yield $y = 4/3$. This 
will then establish the result
\begin{equation}
\label{renor} 
\tilde{\nu} \sim k^{-4/3}, \qquad \tilde{\mu} \sim k^{-4/3},
\qquad \tilde{c}_s \sim k^{-1/3}
\end{equation}
which, it may be mentioned at this point, is identical to the scaling 
relation obtained by \citet{syko90} for the case of a randomly stirred
compressible fluid. It must also be emphasized here that simple dimensional
arguments would not entirely suffice. Indeed, 
\citet{syko90} make this point amply clear by saying that the 
renormalization of the speed of sound is essential to understanding the
physics of compressible flows, since the appearance of the speed of sound
as a dimensional parameter, makes simple dimensional considerations invalid. 

Of immediate interest would be the scaling behaviour of the speed of
sound, which in terms of the radial distance may be written as 
$\tilde{c}_s \sim r^{1/3}$. The steady state solution of the continuity 
equation (with $\rho_0$ written in terms of $c_s$), gives a dependence 
for the steady velocity of the flow $v_0$, which goes as \citep{skc90}
\begin{equation}
\label{acoscon}
v_0 \sim r^{-2} {c_s}^{-2n} 
\end{equation}
where $n=(\gamma -1)^{-1}$ is the polytropic index, whose admissible range 
of values for inflow solutions is given by $3/2 < n < \infty$ \citep{skc90}. 
Using the renormalized speed of sound and its associated scaling relation 
in equation (\ref{acoscon}), will give a scaling behaviour for the steady 
flow velocity as 
\begin{equation}
\label{scalvel}
v_0 \sim r^{-2(1 + n/3)}
\end{equation} 
from which it is quite evident that regardless of the value of $n$, on 
large length scales, the steady flow velocity would die out --- a fact 
that is in conformity with the boundary condition of the flow. The result
given by equation (\ref{scalvel}) highlights another very interesting issue.
It has been discussed earlier that on large length scales, the mean flow
is limited by the equation of continuity, and therefore its variation is
given by $v_0 \sim r^{-2}$. This is a result that is easily derived from the 
classical and inviscid Bondi flow \citep{pso80, skc90}. 
What equation (\ref{scalvel}) 
indicates is that turbulent fluctuations, sustaining themselves at the expense 
of the mean flow, detracts even further from the $r^{-2}$ scaling law for 
the mean flow velocity --- something that, from considerations of energy 
dissipation, can be qualitatively intuited about the influence of turbulence 
on the mean flow. 

The results in equations (\ref{renor}) and (\ref{scalvel}) also lead to the 
conclusion that in the renormalized situation, there would be a scale 
dependence for the position of the sonic horizon as well. For large length 
scales, i.e. concomitantly for a large effective turbulent viscosity, the 
sonic horizon would be shifted inwards. This happens because, seen on a 
large length scale, an enhanced scale dependent speed of sound, could only 
be matched by the steady flow velocity deeper within the gravitational
potential well. Since the flow has to pass through the subsonic region in
any case, this effect of subsonic turbulence in shifting the sonic point
inwards, is also seen to have a bearing on the transonicity of the inflow
solution. This observation would be entirely compatible with
the role of a weak molecular viscosity in inwardly shifting the position of 
the critical point of the inflow solution \citep{ray03}. 

\section{Concluding remarks}

It has been seen so far, how on the large length scales of a spherically
symmetric accreting system, turbulence is capable of setting a scaling
behaviour for both viscosity and the speed of sound. However, it need not
be supposed that given the scaling relation $\tilde{c}_s \sim r^{1/3}$, 
there would be an arbitrarily large scaling for the speed of sound on large 
length scales. This is because in spherical symmetry, turbulence itself 
will also play a role in limiting the accretion process. The physical
quantity ${\dot{m}}/{\nu \rho}$ (with ${\dot{m}}$ being the accretion flow
rate) has the dimension of length, and this has been understood to be a 
viscous shielding radius, $r_{\rm{visc}}$ \citep{ray03}. If the value of 
$\nu$ is enhanced by the introduction of a large and scale dependent  
kinematic viscosity, then $r_{\rm{visc}}$ will define a noticeable spatial 
limit for the accretion process. 

The $r^{1/3}$ scaling behaviour for sound propagation is also apparently 
surprising, with its physical implication being that the flow is heated up 
more at larger radial distances. On the other hand, the classical Bondi
theory shows that the speed of sound increases as the flow moves inward, 
i.e. the flow gets heated up more at smaller radii. The point to remember
is that this property of classical spherical accretion is not violated by 
the mean flow, and the dressing of sound propagation 
that the turbulent fluctuations bring about, is manifest over and above
the standard features that the mean flow is expected to show. The extent
of energy dissipation that turbulence brings about is not accounted for
by the Bondi theory. This energy dissipation shows itself as an enhanced
scaling for the speed of sound (larger scales are more energetic in this
sense), and had temperature been chosen as a 
dynamical variable, this would have shown no contradiction. 
A cautionary reminder that is to be sounded here is that all the scaling 
relations have been derived under the assumption of near incompressibility
on large length scales, which is a condition that cannot be applied too far 
into the inner region of the flow, and in consequence, the $r^{1/3}$ scaling 
for sound propagation is not to be extended too much to small length scales 
either. 

It would also be instructive here to have an understanding of the dynamic
scaling of both the speed of sound and the steady flow velocity, which could 
be derived on using the prescription for an effective viscosity forwarded 
by \citet{ms77}. They proposed a scaling behaviour for the kinematic 
coefficient of turbulent viscosity $\nu_{\mathrm{t}}$, which could be 
conceived of as a product of a characteristic length scale $l_{\mathrm{t}}$ 
of the turbulent cells, and the magnitude of their associated turbulent 
velocity fluctuations $v_{\mathrm{t}}$. It was assumed by \citet{ms77} that 
for all $r$, the length $l_{\mathrm{t}}$ would be some fraction of the radial 
distance $r$, while $v_{\mathrm{t}}$ would be a fraction of the free fall 
velocity $v_{\mathrm{ff}}$, which varies as $r^{-1/2}$. 
For large length scales, 
in which the density of the accreting fluid approaches its constant ambient 
value, this prescription would lead to a scaling behaviour given by 
$\nu_{\mathrm{t}} \sim v_{\mathrm{t}} l_{\mathrm{t}} \sim r^{1/2}$.

In this case, it would then be easy to see 
from the dispersion relation given by equation (\ref{disp}), that 
the speed of sound would be scaled by the relation $c_s \sim r^{-1/2}$, 
while scaling for the steady flow velocity, from equation (\ref{acoscon}), 
would be given by $v_0 \sim r^{n-2}$. The difficulty arises for $n>2$, since 
for large length scales, $v_0$ would actually increase, contrary to a common 
understanding of the boundary condition that $v_0$ should decrease over large 
radial distances. This discrepancy arises because of 
considering the characteristic
eddy velocity to be a fraction of the free fall velocity. 
Even though this looks
well founded on dimensional principles alone, this scaling behaviour 
breaks down
on large length scales, because on these scales free fall 
conditions do not hold.
Rather, this is the region of the {\em ambient} conditions, 
where the mean flow
velocity, even under inviscid conditions, varies at the most 
as $r^{-2}$, and 
therefore the velocity fluctuations
would have to have a different scaling behaviour. Indeed, by being coupled to 
the mean flow, the velocity fluctuations alter the scaling behaviour for the
mean velocity as well, as equation (\ref{scalvel}) indicates. Thus it would 
probably be more correct to suggest that within the sonic radius and close
to the accretor, for a highly supersonic mean flow, free fall conditions can 
have a bearing on the velocity fluctuations. However, the extent of 
the influence 
of turbulence on such small scales would be a somewhat contentious issue, 
and is not within the scope of this work.   

\acknowledgments
This research has made use of NASA's Astrophysics Data System.
One of the authors (AKR) gratefully acknowledges the support provided by
the Council of Scientific and
Industrial Research, Government of India, for a part of the time
that was needed to carry out this work.

\appendix

\section[]{Appendix}

It has been seen that equation (\ref{traneqinu}) is in the form 
\begin{equation}
\label{a1}
\Big[ (- i \omega + \nu k^2) \delta_{ij} + \mu k_i k_j \Big] u_j
- c_s^2 \frac{k_i k_j}{i \omega} u_j = f_i - i \sum_{{\bf{p}},
\omega^{\prime}} p_j u_j ({\bf{k}} - {\bf{p}}, {\omega}^{\prime})
u_i ({\bf{p}}, \omega - {\omega}^{\prime})
\end{equation}
of which, the right hand side is averaged over the distribution of
the random force $f_i$. For the nonlinear term, the perturbative 
expansion of $u_i$ can be written for its first two terms as 
\begin{eqnarray}
\label{a2}
\langle  - i \sum_{{\bf{p}},
\omega^{\prime}} p_j u_j ({\bf{k}} - {\bf{p}},\omega^{\prime})
u_i ({\bf{p}}, \omega - \omega^{\prime}) \rangle 
&=& \langle  - i \sum_{{\bf{p}},
\omega^{\prime}} p_j u_j^{(0)} ({\bf{k}} - {\bf{p}},\omega^{\prime})
u_i^{(0)} ({\bf{p}}, \omega - \omega^{\prime}) \rangle \nonumber \\
& & + \langle  - i \sum_{{\bf{p}}, \omega^{\prime}} p_j 
\big[ u_j^{(1)} ({\bf{k}} - {\bf{p}},\omega^{\prime})
u_i^{(0)} ({\bf{p}}, \omega - \omega^{\prime}) \nonumber \\ 
& & \qquad \qquad +u_j^{(0)} ({\bf{k}} - {\bf{p}},
\omega^{\prime})
u_i^{(1)} ({\bf{p}}, \omega - \omega^{\prime})
\big] \rangle \nonumber \\
\end{eqnarray}

In the above equation, the first term on the right hand, with $u_i^{(0)}$
substituted from equation (\ref{lowodd}), can be written as 
\begin{eqnarray*}
\langle  - i \sum_{{\bf{p}},
\omega^{\prime}} p_j u_j^{(0)} ({\bf{k}} - {\bf{p}},\omega^{\prime})
u_i^{(0)} ({\bf{p}}, \omega - \omega^{\prime}) \rangle &=&
 - i \sum_{{\bf{p}},\omega^{\prime}} p_j G_{jm}^{(0)} 
({\bf{k}} - {\bf{p}},\omega^{\prime}) G_{in}^{(0)}
({\bf{p}}, \omega - \omega^{\prime}) \nonumber \\
& & \qquad \qquad \times \langle f_m({\bf{k}} - {\bf{p}},\omega^{\prime}) 
f_n({\bf{p}}, \omega - \omega^{\prime}) \rangle \nonumber \\
&=&  - i \sum_{{\bf{p}},\omega^{\prime}} p_j G_{jn}^{(0)}
(-{\bf{p}},\omega^{\prime}) C_0(-{\bf{p}}) G_{in}^{(0)}
({\bf{p}},-\omega^{\prime})
\end{eqnarray*}

As can easily be seen, the expression above does not 
produce any momentum $({\bf{k}})$ or frequency $(\omega)$ dependent term and
hence is not responsible for momentum transfer. The second 
term in equation (\ref{a2}), with $u_i^{(1)}$ substituted 
from equation (\ref{solfirst}), can be written down as 
\begin{eqnarray*}
& & \langle  - i \sum_{{\bf{p}},
\omega^{\prime}} p_j \big[
u_j^{(1)} ({\bf{k}} - {\bf{p}},\omega^{\prime})
u_i^{(0)} ({\bf{p}}, \omega - \omega^{\prime})
+u_j^{(0)} ({\bf{k}} - {\bf{p}},\omega^{\prime})
u_i^{(1)} ({\bf{p}}, \omega - \omega^{\prime})
\big] \rangle \nonumber \\
&=& - \langle \sum_{{\bf{p}},\omega^{\prime}} p_j \Big[ G_{jl}^{(0)}
({\bf{k}}-{\bf{p}},\omega^{\prime})\sum_{{\bf{q}},\omega^{\prime \prime}}
q_k u_k^{(0)} ({\bf{k}}-{\bf{p}}-{\bf{q}},\omega^{\prime \prime})
u_l^{(0)} ({\bf{q}},\omega^{\prime}-\omega^{\prime \prime})
u_i^{(0)} ({\bf{p}},\omega -\omega^{\prime}) \nonumber \\
& & + u_j^{(0)} ({\bf{k}}-{\bf{p}},\omega^{\prime}) G_{il}^{(0)}
({\bf{p}}, \omega -\omega^{\prime})\sum_{{\bf{q}},\omega^{\prime \prime}}
q_k u_k^{(0)}({\bf{p}}-{\bf{q}},\omega^{\prime \prime})
u_l^{(0)} ({\bf{q}},\omega -\omega^{\prime}-\omega^{\prime \prime})
\Big] \rangle \nonumber \\
&=& -2 \sum_{{\bf{p}},{\bf{q}},\omega^{\prime},\omega^{\prime \prime}}
p_j \Big[G_{jl}^{(0)}({\bf{k}}-{\bf{p}},\omega^{\prime}) q_k 
G_{km}^{(0)} ({\bf{k}}-{\bf{p}}-{\bf{q}},\omega^{\prime \prime})
G_{in}^{(0)} ({\bf{p}}, \omega -\omega^{\prime}) \nonumber \\
& & \qquad \qquad \times \langle f_m
({\bf{k}}-{\bf{p}}-{\bf{q}},\omega^{\prime \prime}) f_n({\bf{p}}, \omega
-\omega^{\prime}) \rangle u_l^{(0)} ({\bf{q}}, \omega^{\prime} -
\omega^{\prime \prime}) \nonumber \\
& & + G_{il}^{(0)} ({\bf{p}}, \omega - \omega^{\prime}) 
q_k G_{km}^{(0)} ({\bf{p}}-{\bf{q}},\omega^{\prime \prime})
G_{jn}^{(0)} ({\bf{k}}-{\bf{p}}, \omega^{\prime}) \nonumber \\
& & \qquad \qquad \times \langle f_m({\bf{p}}
-{\bf{q}},\omega^{\prime \prime})f_n({\bf{k}}-{\bf{p}},\omega^{\prime})
\rangle u_l^{(0)} ({\bf{q}}, \omega -\omega^{\prime}-\omega^{\prime
\prime}) \Big]
\end{eqnarray*}

The factor of $2$ appears in the expression above because $u_i^{(0)}$
could be expressed in two ways with the help of equation (\ref{lowodd}). 
Using the correlation function implied by equation (\ref{corr}), will 
now give from the result above

\begin{eqnarray*}
& &
-2 \sum_{{\bf{p}},\omega^{\prime}} p_j k_k \Big[ G_{jl}^{(0)}({\bf{k}}
-{\bf{p}}, \omega^{\prime})G_{kn}^{(0)}(-{\bf{p}},- \omega +
\omega^{\prime})G_{in}^{(0)}({\bf{p}}, \omega -\omega^{\prime})
C_0({\bf{p}}) \nonumber \\
& & + G_{il}^{(0)}({\bf{p}}, \omega -\omega^{\prime})
G_{kn}^{(0)}({\bf{p}} -{\bf{k}}, -\omega^{\prime})G_{jn}^{(0)}
({\bf{k}}-{\bf{p}},\omega^{\prime})C_0({\bf{k}}-{\bf{p}})
\Big]u_l^{(0)} ({\bf{k}}, \omega) \nonumber \\
&=& -2 \sum_{{\bf{p}},\omega^{\prime}} p_j k_k \Big[G_{jl}^{(0)}({\bf{k}}
-{\bf{p}}, \omega^{\prime}) \tilde{C}_{ki}^{(0)}({\bf{p}}, \omega
-\omega^{\prime}) \nonumber \\
& & \qquad \qquad + G_{il}^{(0)}({\bf{p}}, \omega -\omega^{\prime})
\tilde{C}_{kj}^{(0)}({\bf{k}}-{\bf{p}},\omega^{\prime})
\Big]u_l^{(0)} ({\bf{k}}, \omega) \nonumber \\
&=& -\sigma_{il}^{(0)} ({\bf{k}},\omega)u_l^{(0)}({\bf{k}}, \omega)
\end{eqnarray*}
where
\begin{displaymath}
\sigma_{il}^{(0)} ({\bf{k}}, \omega )
= 2 \sum_{{\bf{p}},\omega^{\prime}} p_j k_k \Big[G_{jl}^{(0)}
({\bf{k}}-{\bf{p}},\omega^{\prime})\tilde{C}_{ki}^{(0)}({\bf{p}},
\omega - \omega^{\prime}) + G_{il}^{(0)}
({\bf{p}}, \omega - \omega^{\prime})\tilde{C}_{kj}^{(0)}({\bf{k}}
- {\bf{p}},\omega^{\prime}) \Big]
\end{displaymath}
and 
\begin{displaymath}
\tilde{C}_{kj}^{(0)}({\bf{p}},\omega^{\prime}) =
G_{kl}^{(0)}({\bf{p}},\omega^{\prime})C_0({\bf{p}})
G_{lj}^{(0)}(-{\bf{p}},-\omega^{\prime})
\end{displaymath}

In the lowest order of the perturbation theory, the nonlinear term now has 
an equivalent linearized representation given by $\sigma_{il}^{(0)} 
({\bf{k}}, \omega )u_l^{(0)}({\bf{k}}, \omega )$. The $\sigma_{il}^{(0)}$
that has been obtained is called the self energy and it serves the purpose
of dressing the bare coefficients in the equation of motion. Considering
all the higher order terms, $\sigma_{il}^{(0)}$ will be converted to the
full self energy $\sigma_{il}$. The self energy can be 
compared with equation (\ref{greeninv})
and can be seen to make a contribution to the two coefficients of viscosity
and the speed of sound. Seen in this way it can be written as 
\begin{eqnarray}
\label{a3}
\sigma_{ij}^{(0)} ({\bf{k}}, \omega )
&=& 2 \sum_{{\bf{p}},\omega^{\prime}} p_m k_n \Big[G_{mj}^{(0)}
({\bf{k}}-{\bf{p}},\omega^{\prime})\tilde{C}_{ni}^{(0)}({\bf{p}},
\omega - \omega^{\prime}) \nonumber \\
& & \qquad \qquad + G_{im}^{(0)}
({\bf{p}}, \omega - \omega^{\prime})\tilde{C}_{jn}^{(0)}({\bf{k}}
- {\bf{p}},\omega^{\prime}) \Big]\nonumber \\
&=& k^2 \Big[\sigma_1^{(0)}({\bf{k}, \omega})\delta_{ij}
+\sigma_2^{(0)}({\bf{k}}, \omega) \frac{k_i k_j}{k^2} \Big]
\end{eqnarray}
in which $\sigma_1^{(0)}({\bf{k}}, \omega =0)$ dresses the shear viscosity, 
and $\sigma_2^{(0)}({\bf{k}}, \omega)$ dresses the bulk viscosity and 
the speed of sound.

To understand the effect of renormalization, it would be necessary first to
obtain the Green's function by inversion of the matrix 
implied by equation (\ref{greeninv}).
In this way the Green's function is given as 
\begin{equation}
\label{a4}
G_{ij}^{(0)}({\bf{k}}, \omega) = \frac{1}{-i \omega + {\nu} k^2}
\Bigg[\delta_{ij} - k_i k_j \frac{\big(\mu - c_s^2/ {i \omega} \big)}
{-i \omega + \nu k^2 + k^2  \big(\mu - c_s^2/{i \omega} \big)} \Bigg]
\end{equation}

It is important to check if the incompressible limit is to be correctly 
obtained.
The incompressible limit implies $c_s \longrightarrow \infty$ and in that
limit it is seen that $G_{ij}^{(0)}({\bf{k}},\omega)=P_{ij}({\bf{k}})
[-i\omega + \nu k^2]^{-1}$, where $P_{ij}({\bf{k}}) = \delta_{ij} 
-(k_i k_j)/k^2$, is the projection operator, which is as it should be. 

The renormalization of $\nu$, $\mu$ and $c_s$ converts them into the
renormalized quantities $\tilde{\nu}$, $\tilde{\mu}$ and $\tilde{c}_s$ 
respectively. In 
the event of the two coefficients of viscosity and the speed of sound 
getting renormalized, the fully dressed Green's function is given by
\begin{equation}
\label{a5}
G_{ij}({\bf{k}}, \omega) = \frac{1}{-i \omega + \tilde{\nu} k^2}
\Bigg[\delta_{ij} - k_i k_j\frac{\big(\tilde{\mu} - \tilde{c}_s^2/
{i \omega} \big)}{-i \omega + \tilde{\nu} k^2 + k^2  \big(\tilde{\mu} 
- \tilde{c}_s^2/{i \omega} \big)} \Bigg]
\end{equation}

The poles of the Green's function occur at $\omega = -i \tilde{\nu} k^2$ and 
the roots of $\omega^2 = \tilde{c}_s^2 k^2 - i \omega k^2 
(\tilde{\nu} + \tilde{\mu})$, and on solving the quadratic in $\omega$, 
the dispersion relation is
given by 
\begin{equation}
\label{a6}
2\omega = - ik^2(\tilde{\nu} + \tilde{\mu}) \pm 
\sqrt{4\tilde{c}_s^2 k^2 - k^4 (\tilde{\nu} + \tilde{\mu})^2}
\end{equation}

The long wavelength limit ($k$ small) yields
\begin{equation}
\label{a7}
\omega \cong \pm \tilde{c}_s k - \frac{i}{2} k^2 (\tilde{\nu} 
+ \tilde{\mu})
\end{equation}

Dynamic scaling would imply that the frequency would be proportional 
to some power of $k$, and this in its turn would mean that each term 
in the right hand side of equation (\ref{a7}) must scale in the same way.  
Assuming a power law of the form $\tilde{\nu} \propto k^{-y}$, will also
clearly lead to having $\tilde{\mu} \propto k^{-y}$ and 
$\tilde{c}_s \propto k^{1 -y}$. To know the value of $y$, the function 
$C_0 \big( \vert {\bf{r}} - {\bf{r}^{\prime}} \vert \big)$ 
in equation (\ref{corr}) would have to be specified first. Assuming a 
scaling form given by
$C_0 \big( \vert {\bf{r}} - {\bf{r}^{\prime}} \vert \big)
\propto {\vert {\bf{r}} - {\bf{r}^{\prime}} \vert}^{\alpha}$, 
yields the corresponding transformation in the momentum space as 
$C_0 (k) \sim k^{-(D+ \alpha)}$, 
with $D$ being the dimensionality of the space. 

The fully dressed self-consistent form of equation (\ref{a3}) can be set 
in an integral form given by (on noting that the two summations in the 
right hand side of equation (\ref{a3}) are quite identical) 
\begin{equation}
\label{a8}
\sigma_{ij}({\bf{k}}, \omega) = 4 \int \frac{\mathrm{d}^D p}{(2\pi)^D}
\frac{\mathrm{d} \omega^{\prime}}{2\pi} p_m k_n G_{mj}({\bf{k}} - {\bf{p}},
\omega - \omega^{\prime}) \tilde{C}_{ni}({\bf{p}}, \omega^{\prime})
\end{equation}
where $\tilde{C}_{ni} = G_{nr} C_0 G_{ri}^{\ast}$. This is the 
generalization of the self-consistent mode coupling of the incompressible
turbulent flow, to the compressible turbulent flow and is the mode coupling 
version of the renormalization group arguments of \citet{syko90}. As in 
all such problems, the mode coupling integral is valid over a larger 
momentum scale and hence, in principle, allows more than the asymptotic 
analysis of the renormalization group. In this work, the focus is only on 
the exponent $\alpha$, which is an asymptotic result. In the incompressible 
limit, it is to be noted that 
$G_{ij}=P_{ij}(-i \omega + \tilde{\nu} k^2)^{-1}$, 
which forces 
$\tilde{C}_{ij}=P_{ij} \tilde{C}_0 (\omega^2 + \tilde{\nu}^2 k^4)^{-1}
k^{-(D+ \alpha)}$, where $\tilde{C}_0$ is a constant. 
This reduces equation (\ref{a8}) to 
\begin{equation}
\label{a9}
\tilde{\nu}({\bf{k}}, \omega)= \frac{4}{k^2} \int \frac{\mathrm{d}^D {\bf{p}}}
{(2\pi)^D} \int \frac{\mathrm{d} \omega^{\prime}}{2\pi} \frac{k_m k_n P_{mj}
({\bf{k}}-{\bf{p}}) P_{ni}({\bf{p}}) P_{jl}({\bf{k}})P_{li}({\bf{k}})}
{\big[-i(\omega - \omega^{\prime}) + \tilde{\nu} ({\bf{k}}-{\bf{p}})^2 \big]
\big[{\omega}^2 + \tilde{\nu}^2 p^4 \big] p^{D+ \alpha}}
\end{equation}
which is very close to the expression obtained by \citet{bhat91}.
The Kolmogorov spectrum for incompressible turbulence requires that
$\alpha = 0$, which characterizes the nature of the energy transfer between
the mean and the random flow. This transfer characteristic should be
independent of the speed of sound, and so what holds for 
$c_s \longrightarrow \infty$, should also hold at finite $c_s$. 
Consequently, the forcing function is characterized by $\alpha = 0$
in the nearly incompressible regime that is being studied here. 

A scaling analysis 
of equation (\ref{a8}) can now be carried out. The left hand side
scales as $k^{2-y}$. The right hand side clearly scales as $k^{D+2-y+2}
k^{y-2}k^{-D}k^{2(y-2)} = k^{2y-2}$. For the scaling properties of the
right and the left hand sides to agree, it is necessary to have the 
condition $2-y=2y-2$, which gives $y = 4/3$. This leads to the result
$\tilde{\nu} \sim k^{-4/3}, \, \tilde{\mu} \sim k^{-4/3}$ and 
$\tilde{c}_s \sim k^{-1/3}$.

\end{document}